\documentclass[11pt]{article}
\usepackage[english]{babel}
\usepackage[latin1]{inputenc}
\usepackage{amsmath,amsfonts,amssymb}
 \title{Calibration and Internal no-Regret\\ with Partial Monitoring}

\author{Perchet Vianney\thanks{\'Equipe Combinatoire et Optimisation, FRE 3232 CNRS,
Universit\'e Pierre et Marie Curie - Paris 6, 4 place Jussieu, 75005
Paris. vianney.perchet@normalesup.org}}

\begin{document}
\maketitle
\newcounter{compteur}
\newtheorem{proposition}{Proposition}[section]
\newtheorem{theorem}[proposition]{Theorem}
\newtheorem{lemma}[proposition]{Lemma}
\newtheorem{corollary}[proposition]{Corollary}
\newtheorem{definition}[proposition]{Definition}
\newtheorem{remark}[proposition]{Remark}
\newtheorem{example}[proposition]{Example}

\newcounter{Hypo}
\newtheorem{hypo}[Hypo]{Assumption}

\begin{abstract}
Calibrated strategies can be obtained by performing  strategies that
have no internal regret in some auxiliary game. Such  strategies can
be constructed explicitly with the use of Blackwell's
approachability theorem, in an other auxiliary game. We establish
the converse: a strategy that approaches a convex $B$-set can be
derived from the construction of a calibrated strategy.

We develop these tools in the framework of a game with partial
monitoring, where players do not observe the actions of their
opponents but receive random signals, to define a notion of internal
regret and construct strategies that have no such regret.

\bigskip

\end{abstract}

\section*{Introduction}
Calibration, approachability and regret are three notions widely
used both in game theory and machine learning. There are, at first
glance, no obvious links between them. Indeed, calibration has been
introduced by Dawid~\cite{DawidWellCalibrated} for repeated games of
predictions: at each stage, Nature chooses an outcome $s$ in a
finite set $S$ and Predictor forecasts it  by announcing, stage by
stage, a probability over $S$. A strategy is calibrated if the
empirical distribution of  outcomes on the set of stages where
Predictor made a specific forecast is close to it. Foster and
Vohra~\cite{FosterVohraAsymptoticCalibration} proved the existence
of such strategies. Approachability has been introduced by
Blackwell~\cite{BlackwellAnalogue} in two-person repeated games,
where at each stage the payoff is a vector in $\mathbb{R}^d$: a
player can approach a given set $E \subset \mathbb{R}^d$, if he can
ensure that, after some stage and  with a great probability, the
average payoff will always remain close to $E$. This is possible,
see Blackwell~\cite{BlackwellAnalogue}, as soon as $E$ satisfies
some  geometrical condition (it is then called a $B$-set) and this
gives a full characterization for the special case of convex sets.
No-regret has been introduced by Hannan~\cite{Hannan} for two-person
repeated games with payoffs in $\mathbb{R}$: a player has  no
external regret if  his average payoff could not have been
asymptotically better by knowing in advance the empirical
distribution of moves of the other player. The existence of such
strategies was also proved by Hannan~\cite{Hannan}.

\medskip
Blackwell~\cite{BlackwellControlled} (see also Luce and Raifa~\cite{LuceRaiffa},  A.8.6  and Mertens, Sorin and Zamir~\cite{MSZ}, Exercice 7 p.\ 107) was the first to notice that  the
existence of externally consistent strategies (strategies that have
no external regret) can be proved using his approachability theorem.
As shown by Hart and Mas-Colell~\cite{HartMasColellCorrelatedEquilibrium},
the use of Blackwell's theorem actually gives not only  the existence of
externally consistent strategies but also a construction of
strategies that fulfill a stronger property, called internal consistency: a player
has asymptotically no internal regret, if for each of his actions, he
has no external regret on the set of stages where he played it (as long as this set has a positive density). This more precise  definition of regret has been introduced by Foster and Vohra~\cite{FosterVohraCalibratedLearningCorrelatedEquilibrium} (see also Fudenberg and Levine~\cite{FudenbergLevineConditionalUniversalConsistency}).

Foster and Vohra~\cite{FosterVohraAsymptoticCalibration} (see also
Sorin~\cite{SorinUnpublished} for a shorter proof) constructed a
calibrated strategy by computing, in an auxiliary game, a strategy
with no internal regret. These results are recalled in
section~\ref{sectionfull} and we also refer to Cesa-Bianchi and
Lugosi~\cite{CesaBianchiLugosi} for more complete survey on
sequential prediction and regret.

We provide in section \ref{sectionappro} a kind of converse result
by constructing  an explicit $\varepsilon$-approachability strategy
for a convex $B$-set  through the use of a calibrated strategy, in
some auxiliary game. This last statement proves that the
construction of an approachability strategy of a convex set can be
deduced from the construction of a calibrated strategy, which is
deduced from the construction of an internally consistent strategy,
itself deduced from the construction of an approachability strategy.
So calibration, regret and approachability are, in some sense,
\textsl{equivalent}.

\medskip

In section \ref{sectioninfoimparfaite}, we consider repeated games
with partial monitoring, i.e.\  where players do not observe the
action of their opponents, but  receive random signals. The idea
behind the proof that, in the full monitoring case, approachability
follows from calibration can be extended to this new framework to
construct consistent strategies in the following sense. A player has
asymptotically no external regret if his average payoff could not
have been better by knowing in advance the empirical distribution of
signals (see Rustichini~\cite{Rustichini}). The existence of
strategies with no external regret was proved by
Rustichini~\cite{Rustichini} while  Lugosi, Mannor and
Stoltz~\cite{LugosiMannorStoltz} constructed explicitly such
strategies. The notion of internal regret was introduced by Lehrer
and Solan~\cite{LehrerSolanPSE} and they proved the existence of
consistent strategies. Our main result is the construction of such
strategies even when the signal depends on the action played. We
show in section~\ref{sectionexample} that our algorithm also works
when the opponent is not restricted to a finite number of actions,
discuss our assumption on the regularity of the payoff function (see
Assumption \ref{hypo01}) and extend our framework to more general
cases.

\section{Full monitoring case: from approachability to calibration}\label{sectionfull}
We recall the main results about calibration of Foster and Vohra
\cite{FosterVohraAsymptoticCalibration}, approachability of
Blackwell~\cite{BlackwellAnalogue} and regret of Hart and
Mas-Colell~\cite{HartMasColellCorrelatedEquilibrium}. We will prove
some of these results in detail, since they give the main ideas
about the construction of strategies in the partial monitoring
framework, given in section \ref{sectioninfoimparfaite}.
\subsection{Calibration}\label{sectioncalib2} We consider a
two-person repeated game where, at stage $n \in \mathbb{N}$, Nature
(Player~2) chooses an outcome $s_n$ in a finite set $S$ and
Predictor (Player~1) forecasts it by choosing $\mu_n$ in
$\Delta(S)$, the set of probabilities over $S$. We assume
furthermore that $\mu_n$ belongs to a finite set
$\mathcal{M}=\{\mu(l),l\in L\}$. The prediction at stage $n$ is then
the choice of an element $l_n \in L$, called the \textsl{type of
that stage}. The choices of $l_n$ and $s_n$ depend on the past
observations $h_{n-1}=(l_1,s_1,\dots,l_{n-1},s_{n-1})$ and may be
random. Explicitly, the set of finite histories is denoted by $H =
\bigcup_{n \in \mathbb{N}} \left(L \times S\right)^n$, with $\left(L
\times S\right)^0=\emptyset$ and a behavioral strategy $\sigma$ of
Player~1  is a mapping from $H$ to $\Delta(L)$. Given a finite
history $h_n \in \left(L\times S\right)^n$, $\sigma(h_n)$ is the law
of $l_{n+1}$. A strategy $\tau$ of Nature is defined similarly as a
mapping from $H$ to $\Delta(S)$. A couple of strategies
$(\sigma,\tau)$ generates a probability, denoted by
$\mathbb{P}_{\sigma,\tau}$, over $\mathcal{H}=\left(L \times
S\right)^{\mathbb{N}}$, the set of plays endowed with the cylinder
$\sigma$-field.

\medskip
We will use the following notations. For any families
$\mathbf{a}=\{a_m \in \mathbb{R}^d\}_{m \in \mathbb{N}}$ and
$\mathbf{l}=\{l_m \in L \}_{m \in \mathbb{N}}$ and any integer $n
\in \mathbb{N}$, $N_n(l)=\{ 1 \leq m \leq n, l_m =l\}$ is the set of
stages of type $l$ (before the $n$-th), $\overline{a}_n(l)
=\frac{1}{N_n(l)}\sum_{m \in N_n(l)} a_m$ is the average of
$\mathbf{a}$ on this set and $\overline{a}_n=\frac{1}{n}\sum_{m
=1}^n a_m$ is the average of $\mathbf{a}$ over the $n$ first stages.
\begin{definition}[Dawid~\cite{DawidWellCalibrated}]\label{defcalib2}
A strategy $\sigma$ of Player~1 is calibrated with respect to
$\mathcal{M}$ if for every $l \in L$ and every strategy $\tau$ of
Player~2:
\begin{equation}\label{defcalib2equa} \limsup_{n \to +\infty}\frac{|N_n(l)|}{n}\bigg(\|\overline{s}_n(l)-\mu(l)\|_2^2-\|\overline{s}_n(l)-\mu(k)\|_2^2\bigg)\leq0, \quad \forall k \in L, \mathbb{P}_{\sigma,\tau}\mbox{-as},\end{equation}
where $\Delta(S)$ is seen as a subset of $\mathbb{R}^{|S|}$.
\end{definition}
In words, a strategy of Player~1 is calibrated with respect to
$\mathcal{M}$ if $\overline{s}_n(l)$, the empirical distribution of
outcomes when $\mu(l)$ was predicted, is asymptotically closer to
$\mu(l)$ than to any other $\mu(k)$ (or conversely, that $\mu(l)$ is
the closest possible prediction to $\overline{s}_n(l)$), as long as
$|N_n(l)|/n$, the frequency of $l$,  does not go to 0. Foster and
Vohra~\cite{FosterVohraAsymptoticCalibration} proved the existence of such strategies
with an algorithm based on the Expected Brier Score.

\bigskip

An alternative (and more general) way  of defining calibration is
the following. Player~1 is not restricted to make prediction in a finite set
$\mathcal{M}$ and, at each stage, he can choose any probability in
$\Delta(S)$. Consider any finite partition $\mathcal{P}=\{P(k), k \in K\}$ of
$\Delta(S)$ with a diameter small  enough (we recall that the diameter of a partition is defined as $\max_{k\in K} \max_{x,y \in
P(k)} \|x-y\|$). A strategy is $\varepsilon$-calibrated if  the empirical distribution of outcomes
(denoted by $\overline{s}_n(k)$) when the
prediction is in $P(k)$ is  asymptotically $\varepsilon$-close to
$P(k)$ (as long as the frequency of $k \in K$ does not go to zero).
Formally:
\begin{definition}\label{defcalib1}
A strategy $\sigma$ of Player~1 is $\varepsilon$-calibrated if there
exists $\overline{\eta}>0$ such that for every finite partition
$\mathcal{P}=\{P(k),k\in K\}$ of $\Delta(S)$ with diameter smaller
than $\overline{\eta}$ and every strategy $\tau$ of Player~2:
\begin{equation}\label{defcalib1equa} \limsup_{n \to +\infty}\frac{|N_n(k)|}{n}\bigg(d^2\big(\overline{s}_n(k), P(k)\big)-\varepsilon^2\bigg)\leq0, \quad \forall k \in k, \mathbb{P}_{\sigma,\tau}\mbox{-as},\end{equation}
where for every set $E \subset \mathbb{R}^d$ and $z \in
\mathbb{R}^d$, $d(z,E)=\inf_{e \in E}\|z-e\|_2$.
\end{definition}
The following Lemma \ref{lemmacalib} states a calibrated strategy with respect to a grid (as in  Definition \ref{defcalib2})
is $\varepsilon$-calibrated (as in Definition \ref{defcalib1}), therefore we will only use the
first formulation.
\begin{lemma}\label{lemmacalib}
For every $\varepsilon>0$,  there exists  a finite set
$\mathcal{M}=\{\mu(l), l \in L\}$ such that any  calibrated strategy
with respect to $\mathcal{M}$ is $\varepsilon$-calibrated.
\end{lemma}
\textbf{Proof: } Let $\mathcal{M}=\{\mu(l),l\in L\}$ be a finite
$\varepsilon$-grid of $\Delta(S)$: for every probability $\mu \in
\Delta(S)$, there exists $\mu(l) \in \mathcal{M}$ such that $\left\|
\mu-\mu(l)\right\| \leq \varepsilon$. In particular, for every $l
\in L$ and $n \in \mathbb{N}$, there exists $l' \in L$ such that
$\|\overline{s}_n(l)-\mu(l')\|\leq \varepsilon$. Equation
(\ref{defcalib2equa}) implies then that
\[\limsup_{n \to \infty} \frac{|N_n(l)|}{n}\left(d^2(\overline{s}_n(l),\mu(l))-\varepsilon^2\right) \leq 0, \mathbb{P}_{\sigma,\tau}\mbox{-as}.\]
Let $2\overline{\eta}$ be the smallest distance between any two
different $\mu(l)$ and $\mu(l')$. In any  finite partition $\mathcal{P}=\{P(k), k \in K\}$
 of $\Delta(S)$ of diameter smaller
$\overline{\eta}$, $\mu(l)$ belongs to at most one $P(k)$. Hence
$\sigma$ is obviously $\varepsilon$-calibrated.$\hfill \Box$

\begin{remark}
Lemma \ref{lemmacalib} implies that one can  construct an
$\varepsilon$-calibrated strategy as soon as he can construct a
calibrated strategy with respect to a finite $\varepsilon$-grid of
$\Delta(S)$. The size of this grid is in the order of
$\varepsilon^{-|S|}$ (exponential in $\varepsilon$) and it is not
known yet if  there exists an efficient algorithm (polynomial in
$\varepsilon$) to compute $\varepsilon$-calibration. The results
holds with condition (\ref{defcalib1equa}) replaced by \[\limsup_{n
\to +\infty}\frac{|N_n(k)|}{n}\bigg(d\big(\overline{s}_n(k),
P(k)\big)-\varepsilon\bigg)\leq0, \quad \forall k \in k,
\mathbb{P}_{\sigma,\tau}\mbox{-as}\] however Lemma \ref{lemmacalib}
is trivially true with the square terms
$d^2(\overline{s}_n(k),P(k))$ and $\varepsilon^2$.
\end{remark}

\subsection{Approachability}\label{sectionapproch} We will prove in
section \ref{approachtoregret}  that calibration  follows from
no-regret and that no-regret follows from approachability (proofs
originally due to, respectively, Foster and
Vohra~\cite{FosterVohraAsymptoticCalibration} and Hart and
Mas-Colell~\cite{HartMasColellCorrelatedEquilibrium}). We present
here the notion of approachability introduced by
Blackwell~\cite{BlackwellAnalogue}.

Consider a two-person game repeated  in discrete time with vector
payoffs, where at stage $n \in \mathbb{N}$, Player~1 (resp.
Player~2) chooses the action $i_n \in I$ (resp. $j_n \in J$), where
both $I$ and $J$ are finite. The corresponding vector payoff is
$\rho_n =\rho(i_n,j_n)$ where $\rho$ is a mapping from  $I \times J$
into $\mathbb{R}^d$. As usual, a behavioral strategy $\sigma$ (resp.
$\tau$) of Player~1 (resp. Player~2) is a mapping from  the set of
finite histories $H=\bigcup_{n \in \mathbb{N}}\left(I \times
J\right)^n$  to $\Delta(I)$ (resp. $\Delta(J)$).

\bigskip

For a closed set $E \subset \mathbb{R}^d$ and $\delta\geq0$, we
denote by $E^{\delta}=\{z \in \mathbb{R}^d, d(z,E) \leq \delta\}$
the $\delta$-neighborhood of $E$ and by $\Pi_{E}(z) =\left\{e \in E,
d(z,E)=\| z -e \|\right\}$ the set of closest points to $z$ in $E$.

\begin{definition}\label{defapproach2}
\begin{enumerate}
\item[i)]{A closed  set $E \subset \mathbb{R}^d$ is approachable by Player~1 if for every  $\varepsilon >0$, there exist a strategy  $\sigma$ of Player~1 and $N \in \mathbb{N}$, such that for every strategy  $\tau$ of Player~2 and every $n \geq N$:
\[ \mathbf{E}_{\sigma,\tau}\left[d(\overline{\rho}_n,E)\right]\leq \varepsilon \quad \mbox{and} \quad \mathbb{P}\left(\sup_{n \geq N}d(\overline{\rho}_n,E)\geq \varepsilon \right)\leq
\varepsilon.\]

Such a strategy $\sigma$, independent of $\varepsilon$, is called an
approachability strategy of  $E$.}
\item[ii)]{A set $E$ is excludable by Player~2, if there exists  $\delta>0$ such that the complement of  $E^{\delta}$ is approachable by Player~2.}
\end{enumerate}
\end{definition}
In words, a set $E \subset \mathbb{R}^d$ is approachable by
Player~1, if he has a strategy  such that  the average payoff
converges almost surely to $E$, uniformly with respect to the
strategies of Player~2.

\medskip
Blackwell~\cite{BlackwellAnalogue} noticed that a closed set $E$ that fulfills a purely geometrical condition (see Definition \ref{bset2})
is approachable by Player  1. Before stating it, let us denote by
$P^1(x) =\{\rho(x,y), y \in \Delta(J)\}$, the set of expected
payoffs compatible with $x \in \Delta(I)$ and we define similarly
$P^2(y)$.

\begin{definition}\label{bset2}
A closed subset  $E$  of $\mathbb{R}^d$ is a  $B$-set, if for every
$z \in \mathbb{R}^d$, there exist   $p \in \Pi_E(z)$ and
$x\left(=x(z,p)\right) \in \Delta(I)$  such that the hyperplane
through $p$ and perpendicular to $z-p$  separates $z$ from $P^1(x)$,
or formally:
\begin{equation}\label{blackwellcondition2} \forall z \in \mathbb{R}^d, \exists p \in \Pi_{E}(z),  \exists x \in \Delta(I),   \langle \rho(x,y) - p, z - p\rangle \leq 0, \quad \forall y \in \Delta(J).
\end{equation}
\end{definition}

Informally, from any point $z$ outside $E$ there is a closest point
$p$ and a probability $x\in \Delta(I)$ such that, no matter the
choice of Player~2, the expected payoff and $z$ are on different
sides of the hyperplane  through $p$ and perpendicular to $z-p$. To
be precise, this definition (and the following theorem) does not
require that $J$ is finite: one can assume that Player~2 chooses an
outcome vector $U \in [-1,1]^{|I|}$ so that the expected payoff is
$\rho(x,U)=\langle x, U\rangle$.

\begin{theorem}[Blackwell~\cite{BlackwellAnalogue}]\label{stratapprochparfaite2}
If $E$ is a  $B$-set, then $E$ is approachable by Player~1.
Moreover, the strategy $\sigma$ of Player~1 defined by $\sigma(h_n)
= x(\overline{\rho}_n)$ is such that,  for every strategy $\tau$ of
Player~2:
\begin{equation}
\mathbf{E}_{\sigma,\tau}[d^2_E(\overline{\rho}_n)]\leq \frac{4B}{n}
\quad \mbox{and} \quad \mathbb{P}_{\sigma,\tau}\left(\sup_{n \geq
N}d(\overline{\rho}_n,E)\geq \eta\right)\leq \frac{8B}{\eta^2N},
\end{equation}
with $B=\sup_{i,j} \|\rho(i,j)\|^2$.
\end{theorem}

 In the case of a convex
set $C$, a complete characterization is available:
\begin{corollary}[Blackwell~\cite{BlackwellAnalogue}]\label{repoussable2}
A closed convex set $C \subset \mathbb{R}^d$ is approachable by
Player~1 if and only if:
\begin{equation}\label{eqrepoussable}
P^2(y) \cap C \neq \emptyset, \quad  \forall y \in \Delta(J).
\end{equation}
In particular, a closed convex set $C$ is either  approachable by
Player  1, or excludable by Player~2.
\end{corollary}

\begin{remark}\label{remarkapproachconvex}
Corollary \ref{repoussable2} implies that there are (at least) two
different ways to prove that a convex set is approachable. The first
one, called direct proof, consists in proving that $C$ is a $B$-set
while the second one, called  undirect proof, consists  in proving
that  $C$ is not excludable by Player~2, which reduces to find, for
every $y \in \Delta(J)$,  some $x \in \Delta(I)$ such that
$\rho(x,y) \in C$.
\end{remark}
Consider a two-person repeated game in discrete time where, at stage
$n \in \mathbb{N}$, Player~1 chooses $i_n \in I$ as above and
Player~2 chooses a vector $U_n \in [-1,1]^{c}$ (with $c=|I|$). The
associated payoff is $U_n^{i_n}$, the $i_n$-th coordinate of $U_n$.
The internal regret of the stage is the matrix $R_n=R(i_n,U_n)$,
where $R$ is the mapping from $I \times [-1,1]^{c}$ to
$\mathbb{R}^{c^2}$ defined by:

\[R(i,U)^{(i',j)}=\left\{\begin{array}{cc}
0 & \mbox{if} \quad  i' \neq i\\
U^j-U^i & \mbox{otherwise.}\end{array}\right.\]

With this definition, the average internal regret $\overline{R}_n$
is defined by:
\[\overline{R}_n = \left[\frac{\sum_{m \in N_n(i)}\left(U_m^j-U_m^i\right)}{n}\right]_{i,j \in I}=\left[\frac{|N_n(i)|}{n}\left(\overline{U}_n(i)^j-\overline{U}_n(i)^i\right)_{j \in I}\right]_{i \in I}.\]

\begin{definition}[Foster and Vohra~\cite{FosterVohraCalibratedLearningCorrelatedEquilibrium}]\label{definternconst}
A strategy $\sigma$ of Player~1 is internally consistent if for any
strategy $\tau$ of Player~2:
\[\limsup_{n \to \infty} \overline{R}_n \leq 0, \quad \mathbb{P}_{\sigma,\tau}\mbox{-as}.\]
\end{definition}

In words, a strategy is internally
consistent if for every $i \in I$ (with a positive frequency), Player~1 could not have increased
his payoff if he had known, before the beginning of the game, the
empirical distribution of Player~2's actions on $N_n(i)$. Stated
differently, when Player~1 played action $i$, it was his best
(stationary) strategy.
The existence of such strategies have been first proved by Foster and
Vohra~\cite{FosterVohraCalibratedLearningCorrelatedEquilibrium} and Fudenberg and Levine
\cite{FudenbergLevineConditionalUniversalConsistency}.

\begin{theorem}\label{theointern}
There exist internally consistent strategies.
\end{theorem}

Hart and Mas-Colell~\cite{HartMasColellCorrelatedEquilibrium} noted
that an internally consistent strategy can be obtained by
constructing a strategy that approaches the negative orthant
$\Omega=\mathbb{R}^{c^2}_-$ in the auxiliary game where the vector
payoff at stage $n$ is $R_n$. Such a strategy, derived from
approachability theory, is  stronger than just internally consistent
since the regret converges to the negative orthant uniformly with
respect to Player~2's strategy (which was not required in Definition
\ref{definternconst}).

The proof  of the fact that  $\Omega$ is a $B$-set relies on the two
followings lemmas: Lemma \ref{geometric} gives a geometrical property
of $\Omega$ and Lemma \ref{lemmapayoff} gives a property of the
function $R$.

\subsection{From approachability to internal no-regret}\label{approachtoregret}
\begin{lemma}\label{geometric} Let $\Pi_{\Omega}(\cdot)$ be the projection onto $\Omega$. Then, for
every $A \in \mathbb{R}^{c^2}$:
\begin{equation}\label{BSET1}
\left\langle \Pi_{\Omega}(A),A-\Pi_{\Omega}(A)\right\rangle =0.
\end{equation}
\end{lemma}
\textbf{Proof: }   Note that since $\Omega=\mathbb{R}^{c^2}_-$ then
$A^+=A-\Pi_{\Omega}(A)$ where $A^+_{ij}=\max\left(A_{ij},0\right)$
and similarly $A^-= \Pi_{\Omega}(A)$.
 The result is just a rewriting of $\left\langle A^-,A^+\right\rangle=0$. $\hfill \Box$

For every $(c\times c)$-matrix $A =(a_{ij})_{i,j\in I}$ with
non-negative coefficients, $\lambda \in \Delta(I)$ is an invariant
probability of $A$ if for every $i \in I$:
\[\sum_{j\in I}\lambda(j)a_{ji}=\lambda(i)\sum_{j \in I}a_{ij}.\]
The existence of an invariant probability follows from the similar
result for Markov chains,  implied by Perron-Frobenius Theorem (see
e.g.\ Seneta \cite{Seneta}).

\begin{lemma}\label{lemmapayoff}
Let $A =(a_{ij})_{i,j\in I}$ be a non-negative matrix. Then for
every $\lambda$, invariant probability of $A$,  and every $U \in
\mathbb{R}^{c}$:
\begin{equation}\label{BSET2}
\left\langle
A,\mathbf{E}_{\lambda}\left[R(\cdot,U)\right]\right\rangle=0.
\end{equation}
\end{lemma}

\textbf{Proof: }   The $(i,j)$-th coordinate of
$\mathbf{E}_{\lambda}\left[R(\cdot,U)\right]$ is
$\lambda(i)\left(U^j-U^i\right)$, therefore:
\[ \left\langle A,\mathbf{E}_{\lambda}\left[R(\cdot,U)\right]\right\rangle = \sum_{i,j \in I}a_{ij}\lambda(i)\left(U^j-U^i\right)\]
and the coefficient of each $U^i$ is $\sum_{j \in I}a_{ij}
\lambda(i)-\sum_{j\in I}a_{ji}\lambda(j)=0$, because $\lambda$ is an
invariant measure of $A$. Therefore $\left\langle
A,\mathbf{E}_{\lambda}\left[R(\cdot,U)\right]\right\rangle=0$.
$\hfill \Box$

\textbf{Proof of Theorem \ref{theointern}:} Summing equations
(\ref{BSET1}) (with $A = \overline{R}_n$) and (\ref{BSET2}) (with $A
= \left(\overline{R}_n\right)^+$) gives:
\[ \left\langle \mathbf{E}_{\lambda_n}\left[R(\cdot,U)\right]-\Pi_{\Omega}(\overline{R}_n),\overline{R}_n-\Pi_{\Omega}(\overline{R}_n)\right\rangle=0,\]
for every $\lambda_n$ invariant probability of $\overline{R}_n^+$
and every $U \in [-1,1]^I$.

Define the strategy $\sigma$ of Player~1 by $\sigma(h_n)=\lambda_n$.
The expected payoff at stage $n+1$ (given $h_n$ and $U_{n+1}=U$) is
$\mathbf{E}_{\lambda_n}\left[R(\cdot,U)\right]$, so $\Omega$ is a
$B$-set and is approachable by Player~1. $\hfill \Box$

\begin{remark}
The construction of the strategy is based on  approachability
properties therefore the convergence is uniform with respect to the
strategies of Player~2. Theorem \ref{stratapprochparfaite2} implies
that for every $\eta>0$, and  for every strategy $\tau$ of Player~2:
\[\mathbb{P}_{\sigma,\tau}\left(\exists n \geq N, \exists i,j \in i, \frac{|N_n(i)|}{n}\left(
\overline{U}_n(i)^j-\overline{U}_n(i)^i \right)> \eta\right) =
O\left(\frac{1}{\eta^2N}\right)
\]
\[
\mbox{and} \quad \mathbf{E}_{\sigma,\tau}\left[\sup_{i \in
I}\frac{|N_n(l)|}{n}\left(\overline{U}_n(i)^j-\overline{U}_n(i)^i\right)^+\right]=
O\left(\frac{1}{\sqrt{n}}\right).
\]
Although they are not required by definition \ref{definternconst},
those bounds will be useful to prove that calibration implies
approachability.
\end{remark}

\subsection{From internal regret to calibration}\label{regretocalib}
The construction of calibrated strategies can be reduced to the
construction of internally consistent strategies. The proof  of
Sorin~\cite{SorinUnpublished} simplifies the one originally due to  Foster and
Vohra~\cite{FosterVohraCalibratedLearningCorrelatedEquilibrium} by using the following lemma:

\begin{lemma}\label{lemmamoyenne}
Let $(a_m)_{m \in \mathbb{N}}$ be a sequence in $\mathbb{R}^d$ and
$\alpha$, $\beta$ two points in $\mathbb{R}^d$. Then for every  $n
\in \mathbb{N}^*$:
\begin{equation}\label{equalemmamoyenne}
\frac{\sum_{m=1}^n \left\| a_m-\beta\right\|^2_2-\left\|
a_m-\alpha\right\|^2_2 }{n}=\left\|
\overline{a}_n-\beta\right\|^2_2- \left\|
\overline{a}_n-\alpha\right\|^2_2,
\end{equation}
with  $\| \cdot\|_2$ the Euclidian norm of $\mathbb{R}^d$.
\end{lemma}

\textbf{Proof: } Develop the sums in equation
(\ref{equalemmamoyenne}) to get the result. $\hfill \Box$

Now, we can prove the following:
\begin{theorem}[Foster and Vohra~\cite{FosterVohraCalibratedLearningCorrelatedEquilibrium}]\label{theocalibration}
Let $\mathcal{M}$ be a  finite grid of $\Delta(S)$. There exist
calibrated strategies of Player~1 with respect to $\mathcal{M}$. In
particular, for every $\varepsilon >0$ there exist
$\varepsilon$-calibrated strategies.
\end{theorem}
\textbf{Proof: }   We start with the framework described in section
\ref{sectioncalib2}. Consider the auxiliary two-person game with
vector payoff defined as follows. At stage $n \in \mathbb{N}$,
Player~1 (resp. Player~2) chooses the action $l_n \in L$ (resp. $s_n
\in S$)  which generates the vector payoff $R_n=R(l_n,U_n) \in
\mathbb{R}^{d}$, where $R$ is as in \ref{approachtoregret}, with:
\[ U_n = \left(-\left\| s_n-\mu(l)\right\|^2_2\right)_{l \in L} \in \mathbb{R}^{c}.\]

By definition of $R$ and using Lemma \ref{lemmamoyenne},  for every
$n \in \mathbb{N}^*$:
\begin{eqnarray}\nonumber\overline{R}_n^{lk} &=& \frac{|N_n(l)|}{n}\left(\frac{\sum_{m \in N_n(l)}\left\| s_m-\mu(l)\right\|^2_2-\left\| s_m-\mu(k)\right\|^2_2}{|N_n(l)|}\right)\\
\nonumber&=&\frac{|N_n(l)|}{n}\left(\left\|
\overline{s}_n(l)-\mu(l)\right\|^2_2-\left\|
\overline{s}_n(l)-\mu(k)\right\|^2_2\right).
\end{eqnarray}

Let $\sigma$ be an internally consistent strategy in this auxiliary
game, then for every $l \in L$ and $k \in L$:
\[\limsup_{n \to
\infty}\frac{|N_n(l)|}{n}\left(\left\|\overline{s}_n(l)-\mu(l)\right\|^2_2-\left\|\overline{s}_n(k)-\mu(k)\right\|^2_2\right)
\leq 0, \quad \mathbb{P}_{\sigma,\tau}\mbox{-as}.\] Therefore
$\sigma$ is calibrated, with respect to $\mathcal{M}$; if it  is an
$\varepsilon$-grid of $\Delta(S)$, then $\sigma$ is
$\varepsilon$-calibrated. $\hfill \Box$

\begin{remark}\label{remarkcalib2} We have proved that $\sigma$ is such that,
for every $l\in L$, $\overline{s}_n(l)$ is closer to $\mu(l)$ than
to any other $\mu(k)$, as soon as $|N_n(l)|/n$ is not too small.

The facts that $s_n$ belongs to a finite set $S$ and $\{\mu(l)\}$
are probabilities over $S$ are irrelevant:  one can show that for
any finite set $\{a(l) \in \mathbb{R}^d, l\in L\}$, Player~1 has a
strategy $\sigma$ such that for any bounded sequence $(a_m)_{m \in
\mathbb{N}}$ in $\mathbb{R}^d$ and  for every $l$ and $k$ :
\[\limsup_{n \to \infty} \frac{|N_n(l)|}{n}\bigg(\left\| \overline{a}_n(l)-a(l)\right\|^2-\left\|
\overline{a}_n(l)-a(k)\right\|^2\bigg)\leq0.\]
\end{remark}

\subsection{From calibration to approachability}\label{sectionappro}

The proof of  Theorem \ref{theocalibration} shows that  the
construction of a  calibrated  strategy  can be obtained through an
approachability strategy of an orthant in an auxiliary game.

Conversely, we will show that the approachability of a convex
$B$-set can be reduced to the existence of a calibrated strategy in
an auxiliary game, and so give a new proof of  Corollary
\ref{repoussable2} (and mainly construct explicit strategies).

\medskip

\textbf{Alternative proof of  Corollary \ref{repoussable2}:} The idea
of the proof is very natural: assume that condition~(\ref{eqrepoussable}) is satisfied and rephrased as:
\begin{equation}\label{conditionrephrased2}\forall y \in \Delta(J), \exists x(=x_y) \in \Delta(I), \rho(x_y,y) \in C.\end{equation}
If Player~1 knew in advance $y_n$ then he would just have to play
accordingly to $x_{y_n}$ at stage $n$ so that the expected payoff
$\mathbf{E}_{\sigma,\tau}[\rho_n]$ would be in $C$. Since $C$ is
convex, the average payoff would also be in $C$. Obviously Player~1
does not know $y_n$ but, using calibration, he can make {\em good}
predictions about it.

\medskip

Since $\rho$ is multilinear and therefore continuous on
$\Delta(I)\times\Delta(J)$, for every $\varepsilon>0$, there exists
$\delta>0$ such that:
\[\forall y,y' \in \Delta(J), \left\| y-y'\right\|_2 \leq 2\delta \Rightarrow \rho(x_y,y') \in C^{\varepsilon}.\]
We introduce the auxiliary  game $\Gamma$ where Player~2 chooses an
action  (or outcome) $j \in J$ and Player~1 forecasts it by using
$\{y(l),l\in L\}$, a finite $\delta$-grid of $\Delta(J)$. Let
$\sigma$ be a  calibrated strategy for Player~1, so that
$\overline{\jmath}_n(l)$, the empirical distribution of actions of
Player~2 on $N_n(l)$,  is asymptotically $\delta$-close to $y(l)$.

\medskip
Define the strategy of Player~1 in the initial game by performing
$\sigma$ and if $l_n=l$ by playing accordingly to $x(l):=x_{y(l)}
\in \Delta(I)$, as depicted in (\ref{conditionrephrased2}). Since
the choices of actions of the two players are independent,
$\overline{\rho}_n(l)$ will be close to
$\rho\left(x(l),\overline{\jmath}_n(l)\right)$, hence close to
$\rho(x(l),y(l))$ (because $\sigma$ is calibrated) and finally close
to $C^{\varepsilon}$, as soon as $|N_n(l)|$ is not too small.

\bigskip
Indeed, by construction of $\sigma$, for every $\eta >0$ there
exists $N_1 \in \mathbb{N}$ such that, for every strategy $\tau$ of
Player~2:
\[\mathbb{P}_{\sigma,\tau}\left(\forall l \in L, \forall n \geq N_1, \frac{|N_n(l)|}{n}\left(\left\|\overline{\jmath}_n(l)-y(l)\right\|^2_2 -\delta^2\right)\leq \eta\right)\geq 1-\eta.\]
This implies that with probability greater than $1-\eta$, for every
$l \in L$ and $n \geq N_1$,
 either $\|\overline{\jmath}_n(l)-y(l)\| \leq 2 \delta$ or $N_n(l)/n \leq \eta/3\delta^2$,
 therefore with $\mathbb{P}_{\sigma,\tau}$-probability at least $1- \eta$:
\begin{equation}\label{equaappro12}\forall l \in L, \forall n \geq N_1, \frac{|N_n(l)|}{n}d\left(\rho(x(l),\overline{\jmath}_n(l)),C\right)\leq \varepsilon\frac{|N_n(l)|}{n}+\frac{\eta}{3\delta^2}.\end{equation}

Hoeffding-Azuma~\cite{Azuma,Hoeffding} inequality for sums of bounded martingale differences
 implies that for any $\eta>0$, $n \in
\mathbb{N}$, $\sigma$ and $\tau$:
\[\mathbb{P}_{\sigma,\tau}\left(\left|\overline{\rho}_n(l)-\rho(x(l),\overline{\jmath}_n(l))\right|\geq \eta \big||N_n(l)|\right)\leq2\exp\left(-\frac{|N_n(l)|\eta^2}{2}\right),\]
therefore:
\[\mathbb{P}_{\sigma,\tau}\left(\frac{|N_n(l)|}{n}\left|\overline{\rho}_n(l)-\rho(x(l),\overline{\jmath}_n(l))\right|\geq \eta \right)\leq2\exp\left(-\frac{n\eta^2}{2}\right)\]
and summing over $n \in \{N, \dots,\}$ and $l \in L$ gives that with
$\mathbb{P}_{\sigma,\tau}$-probability at most
$\frac{4L}{\eta^2}\exp\left(-\frac{N\eta^2}{2}\right)$
\begin{equation}\label{equaappro22}\sup_{ n \geq N}\sup_{l \in L} \left\{\frac{|N_n(l)|}{n}\left|\overline{\rho}_n(l)-\rho(x(l),\overline{\jmath}_n(l))\right|\right\}\geq \eta.\end{equation}
So for every $\eta>0$, there exists $N_2 \in \mathbb{N}$ such that
for every $n \geq N_2$:
\[\mathbb{P}_{\sigma,\tau}\bigg(\forall m \geq n, \forall l \in L, \frac{|N_n(l)|}{n}\left|\overline{\rho}_n(l)-\rho(x(l),\overline{\jmath}_n(l))\right| \leq \eta \bigg) \geq 1-\eta.\]

Since $C$ is a convex set, $d(\cdot,C)$ is convex and with
probability at least $1-2\eta$, for every $n \geq \max(N_1,N_2)$:
\begin{align*}d\left(\overline{\rho}_n,C\right)=&d\left(\sum_{l \in L} \frac{|N_n(l)|}{n}\overline{\rho}_n(l),C\right)\leq \sum_{l \in L}\frac{|N_n(l)|}{n}d\left(\overline{\rho}_n(l),C\right)\\ \leq&\sum_{l \in L}\frac{|N_n(l)|}{n}\bigg[d\left(\rho(x(l),\overline{\jmath}_n(l)),C\right)+\left|\overline{\rho}_n(l)-\rho(x(l),\overline{\jmath}_n(l))\right|\bigg]\\
\leq& \varepsilon+L\eta\left(\frac{1}{3\delta^2}+1\right).
 \end{align*}
And $C$ is approachable by Player~1.

\bigskip

On the other hand, if there exists $y$ such that $P^2(y) \cap C =
\emptyset$, then Player~2 can approach $P^2(y)$, by playing at every
stage accordingly to $y$. Therefore $C$ is not approachable by
Player~1. $\hfill \Box$

\begin{remark}
To deduce that $\overline{\rho}_n$ is in $C^{\varepsilon}$ from the
fact that $\overline{\rho}_n(l)$ is in $ C^{\varepsilon}$ for every
$l \in L$, it is necessary that $C$ (or $d(\cdot,C)$) is convex. So
this proof does not work if $C$ is not convex.
\end{remark}

\subsection{Remarks on the algorithm}\label{remarkapproch}
\begin{itemize}
\item[a)]{Blackwell proved Corollary \ref{repoussable2} using Von Neumann's minmax theorem, the latter allowing  to show that a convex set $C$ that fulfills condition
 (\ref{conditionrephrased2}) is a $B$-set.
Indeed, let $z$ be a point outside $C$. Recall that for
every  $y \in \Delta(J)$ there exists $x_y \in \Delta(I)$ such that
$\rho(x_y,y) \in C$. Since $C$ is convex, if we denote by $\Pi_C(z)$
the projection of $z$ onto it, then for every $c \in C$
$\langle c-\Pi_C(z),z-\Pi_C(z)\rangle\leq 0$, . Therefore,
\[\forall y \in \Delta(J), \exists x \in \Delta(J), \langle \mathbf{E}_{x,y}[\rho(i,j)]-\Pi_C(z),z-\Pi_C(z)\rangle \leq0\]
and if we define $g(x,y)=\langle
\mathbf{E}_{x,y}[\rho(i,j)]-\Pi_C(z),z-\Pi_C(z)\rangle$ then $g$ is
linear in both of its variable so \[\min_{x \in \Delta(I)}\max_{y
\in \Delta(J)}g(x,y)=\max_{y \in \Delta(J)}\min_{x \in
\Delta(J)}g(x,y)\leq 0,\] which implies that $C$ is a $B$-set.

The strategy $\sigma$ defined by $\sigma(h_n)=x_n$ where $x_n$ is
any minimizer of $\max_{y \in \Delta(J)}G(x,y)$ is an
approachability strategy, said to be {\em implicit} since  there are
no easy way to construct it. Indeed computing $\sigma$ would require
to find, stage by stage, an optimal action in a zero-sum game or
equivalently to solve a Linear Program. There exist  polynomial
algorithms (see Khachiyan~\cite{Khachiyan}) however their rates of
convergence are bigger than the one of Gaussian elimination and
their constants can be too huge for any practical use. Nonetheless,
it is possible to find $\varepsilon$-optimal solution by repeating
an polynomial number of time the \textsl{exponential weight
algorithm} (see Cesa-Bianchi and Lugosi~\cite{CesaBianchiLugosi},
Section 7.2 and Mannor and Stoltz~\cite{MannorStoltz}).

For  a fixed $\varepsilon>0$, the strategy (that approaches
$C^{\varepsilon}$) we described computes at each stage an invariant
measure of a matrix with non-negative coefficients. This obviously
reduces to solve a system of linear equations which is guaranteed to
have a solution. And this is solved polynomially (in $|L|$) by, for
example and as proposed by Foster and Vohra~\cite{FosterVohraCalibratedLearningCorrelatedEquilibrium}, a
Gaussian elimination. If payoffs are bounded by 1, then one can
 take for
$\{y(l), l\in L\}$ any arbitrarily $\varepsilon/2$-grid of $\Delta(J)$, so $|L|$ is bounded by $(2/\varepsilon)^{|J|}$.
Moreover, the strategy aims to approach $C^{\varepsilon}$, so it is not
compulsory to determine exactly $x(l)$, one can choose them in any
$\varepsilon/2$-grid of $\Delta(I)$.

In conclusion, Blackwell's \textsl{implicit} algorithm constructs a
strategy that approaches (exactly) a convex $C$ by solving, stage by
stage, a Linear Program without any initialization phase. For every
$\varepsilon>0$, our \textsl{explicit} algorithm constructs a
strategy that approaches $C^{\varepsilon}$ by solving, stage by
stage, a system of linear equations with an initialization phase
(the matchings between $x(l)$ and $y(l)$) requiring at most
$(2/\varepsilon)^{I+J}$ steps.

\medskip
}
\item[b)]{Blackwell's Theorem states that if for every move $y \in \Delta(J)$ of Player~2, Player~1 has an action $x \in \Delta(I)$ such that $\rho(x,y) \in C$ then $C$ is approachable by Player~1. In other words, assume that in the one-stage (expected) game where Player~2 plays first and Player~1 plays second, Player~1 has a strategy such that the payoff is in a convex $C$. Then he also has a strategy such that the average payoff converges to $C$, in the repeated (expected) game where Player~2 plays second and Player~1  plays first.

The use of calibration transforms this implicit statement into  an
explicit one: while performing a calibrated strategy (in an
auxiliary game where $J$ plays the role of the set of outcomes),
Player~1 can enforce the property that, for every $l \in L$, the
average move of Player~2 is almost $y(l)$ on $N_n(l)$. So he just
has  to play $x_{y(l)}$ on these stage and he could not do better.
\medskip
}
\item[c)]{
We stress out the fact that  the construction of an approachability
strategy of $C^{\varepsilon}$ reduces to the construction of a
calibrated strategy in an auxiliary game, hence to the
construction of an internally-consistent strategy in a second
auxiliary game, therefore to the construction of an
approachability strategy of a negative orthant in a third auxiliary
game. In conclusion, the approachability of an arbitrary convex set
reduces to the approachability of an orthant. Along
with equations (\ref{equaappro12}) and (\ref{equaappro22}), this  implies that
$\mathbf{E}_{\sigma,\tau}\left[d\left(\overline{\rho}_n,C\right)-\varepsilon\right]\leq
O\left( n^{-1/2}\right)$. However, as said before, the constant
depends on $\varepsilon^{|J|}$.
\medskip
}
\item[d)]{
The reduction of the approachability of a convex set $C \subset
\mathbb{R}^d$ in a game $\Gamma$  to the approachability of an
orthant in an auxiliary game $\Gamma'$ can also be done via the
following scheme: for every $\varepsilon>0$, find a finite set of
half-spaces $\{H(l), l \in L\}$ such that $C \subset \cap_{l \in L}
H(l) \subset C^{\varepsilon}$. For every $l \in L$, define $c(l) \in
\mathbb{R}^d$ and $b(l) \in \mathbb{R}$ such that:
\[H(l)=\left\{\omega \in \mathbb{R}^d, \langle \omega, c(l) \rangle \leq b(l) \right \}\] and the auxiliary game $\Gamma'$ with payoffs defined by
\[\widehat{\rho}(i,j)=\left( \langle\rho(i,j),c(l)\rangle - b(l)\right)_{l \in L} \in \mathbb{R}^L.\]
Obviously, a strategy that approaches the negative orthant in
$\Gamma'$ will approach, in the game $\Gamma$, the set $\bigcap
H(l)$ and therefore $C^{\varepsilon}$. However, such a strategy
might not be based on regret and  might not be explicit.}
\end{itemize}

\section{Internal regret in the partial monitoring framework}\label{sectioninfoimparfaite}
Consider a two person game repeated  in discrete time. At stage $n
\in \mathbb{N}$, Player~1 (resp. Player~2) chooses  $i_n \in I$
(resp. $j_n \in J$), which generates the payoff
$\rho_n=\rho(i_n,j_n)$ where $\rho$ is a mapping from $I \times J$
to $ \mathbb{R}$. Player~1 does not observe this payoff, he receives
a signal $s_n \in S$ whose law is $s(i_n,j_n)$ where $s$ is a
mapping from $I \times J$ to $\Delta(S)$. The three sets $I$, $J$
and $S$ are finite and the two functions $\rho$ and $s$ are extended
to $\Delta(I)\times \Delta(J)$ by
$\rho(x,y)=\mathbb{E}_{x,y}[\rho(i,j)] \in \mathbb{R}$ and
$s(x,y)=\mathbb{E}_{x,y}[s(i,j)] \in \Delta(S)$.

We define the mapping $\mathbf{s}$ from $\Delta(J)$ to $\Delta(S)^I$
by $\mathbf{s}(y)=\left(s(i,y)\right)_{i \in I}$ and we call such a vector of
probability a flag. Player~1 cannot distinguish between  two different probabilities
$y$ and $y'$ in $\Delta(J)$ that induces the same flag $\mu \in \Delta(S)^I$, i.e.\ such that
$\mu=\mathbf{s}(y)=\mathbf{s}(y')$. Thus we say that
$\mu=\mathbf{s}(y)$, although unobserved, is the \textsl{relevant or
maximal} information available to Player~1 about the choice of
Player~2. We stress out that a flag $\mu$ is not observed  since
given $x \in \Delta(I)$ and $y \in \Delta(J)$, Player~1 has just an
information about $\mu^{i}$ which is only one component of $\mu$  (the $i$-th one, where $i$ is the realization of $x$). Moreover, this component is the law of a random variable whose realization (i.e.\ the signal $s \in S$) is the only observation of Player~1.

\begin{example}[Label efficient prediction]\label{examplelabelefficient}

Consider the following game (Example 6.4 in Cesa-Bianchi and Lugosi~\cite{CesaBianchiLugosi}). Nature chooses an outcome $G$ or $B$ and Player~1 can either observe the actual outcome (action $o$) or choose to not observe it and to pick a label $g$ or $b$. If he chooses the right label,  his payoff is  1 and otherwise 0. Payoffs  and laws of signals received by  Player~1 can be resumed by the following matrices (where $a$, $b$ and $c$ are three different probabilities over a finite set $S$).
\begin{center}
\begin{tabular}{cc|c|c|cc|c|c|}
&\multicolumn{1}{c}{}& \multicolumn{1}{c}{$G$}&\multicolumn{1}{c}{$B$}&&\multicolumn{1}{c}{}&\multicolumn{1}{c}{$G$}&\multicolumn{1}{c}{$B$}\\
\cline{3-4}\cline{7-8}
&$o$&0&0&&$o$&$a$&$b$\\
\cline{3-4}\cline{7-8}
Payoffs: &$g$&0&1&\quad and Signals:&$g$&$c$&$c$\\
\cline{3-4}\cline{7-8}
&$b$&1&0&&$b$&$c$&$c$\\
\cline{3-4}\cline{7-8}
\end{tabular}
\end{center}
Action $G$, whose best response is $g$, generates the flag $(a,c,c)$ and  action $B$, whose best response is $b$, generates the flag $(b,c,c)$.  In order to distinguish between those two  actions, Player~1 needs to know the entire flag and therefore to know $s(o,y)$ although  action $o$ is never a best response (but is said to be \textsl{purely informative}).
\end{example}
As usual, a behavioral strategy $\sigma$ of Player~1 (resp. $\tau$
of Player~2) is a function from the set of finite histories for
Player~1, $H^1=\bigcup_{n \in \mathbb{N}}\left(I\times S\right)^n$,
to $\Delta(I)$ (resp. from $H^2=\bigcup_{n \in \mathbb{N}}\left(I
\times S \times J\right)^n$ to $\Delta(J)$). A couple
$(\sigma,\tau)$ generates a probability $\mathbb{P}_{\sigma,\tau}$
over $\mathcal{H}=\left(I \times S \times J\right)^{\mathbb{N}}$.

\subsection{External regret} Rustichini
\cite{Rustichini} defined external consistency in the  partial
monitoring framework as follows: a strategy $\sigma$ of Player~1 has
no external regret if $\mathbb{P}_{\sigma,\tau}$-as:
\[\limsup_{n \to +\infty}\max_{x \in \Delta(I)}\min_{ \left\{\begin{array}{c}y \in \Delta(J),\\ \mathbf{s}(y)=\mathbf{s}(\overline{\jmath}_n)\end{array}\right.}\rho(x,y) -\overline{\rho}_n \leq 0.\]
where $\mathbf{s}(\overline{\jmath}_n)\in \Delta(S)^I$ is the
average flag. In words, the average payoff of Player~1 could not
have been uniformly better  if he had known the average distribution
of flags before the beginning of the game.

\bigskip
Given a flag $\mu \in \Delta(S)^I$, the function $\min_{y \in
\mathbf{s}^{-1}(\mu)}\rho(\cdot,y)$ may not be linear. So the best
response of Player~1 might not be a pure action in $I$, but a mixed
action $x \in \Delta(I)$ and any pure action in the support of $x$
may be a bad response. This explains why, in Rustichini's
definition, the maximum is taken over $\Delta(I)$ and not just over
$I$ as in the usual definition of external regret.

\begin{example}[Matching Penny in the dark] Player~1 chooses  either $T$ail or $H$eads and flips a  coin.
Simultaneously, Nature chooses on which face the coin will land. If
Player~1 guessed correctly his payoff equals 1, otherwise  -1. We
assume  that  Player~1 does not observe the  coin.

Payoffs and signals are resumed in the following matrices:
\begin{center}
\begin{tabular}{cc|c|c|cc|c|c|}
&\multicolumn{1}{c}{}& \multicolumn{1}{c}{$T$}&\multicolumn{1}{c}{$H$}&&\multicolumn{1}{c}{}&\multicolumn{1}{c}{$T$}&\multicolumn{1}{c}{$H$}\\
\cline{3-4}\cline{7-8}
Payoffs:&$T$&1&-1&\quad and Signals:&$T$&$c$&$c$\\
\cline{3-4}\cline{7-8}
&$H$&-1&1&&$H$&$c$&$c$\\
\cline{3-4}\cline{7-8}
\end{tabular}
\end{center}
Every choice of Nature  generates the same flag $(c,c)$. So $\min_{y
\in \Delta(J)}\rho(x,y)$ is always non-positive and equals  zero
only if $x=(1/2,1/2)$. Therefore the only best response of Player~1
is $(1/2,1/2)$, while both $T$ or $H$ give the worst payoff of -1.
\end{example}

\subsection{Internal regret}
We consider here a generalization of the previous's framework: at
stage $n \in \mathbb{N}$, Player~2 chooses a flag $\mu_n \in
\Delta(S)^I$ while Player~1 chooses an action $i_n$ and receives a
signal $s_n$ whose law is the $i_n$-th coordinate of $\mu_n$. Given
a flag $\mu$ and $x \in \Delta(I)$, Player~1 evaluates the payoff
through an evaluation function $G$ from $\Delta(I) \times
\Delta(S)^I$ to $\mathbb{R}$, which is not necessarily linear.

\medskip

Recall that with full monitoring, a strategy has no internal regret
if each action $i \in I$ is the best response to the average
empirical observation on the set of stages where $i$ was actually
played. With partial monitoring,   best responses are elements of
$\Delta(I)$ and not elements of $I$,  so if we want to define
internal regret in this framework, we have to distinguish the stage
not as a function of the action actually played (i.e.\ $i_n \in I$)
but as a function of its law (i.e.\ $x_n \in \Delta(I)$). We assume
that the strategy of Player~1 can be described by a finite family
$\{x(l) \in \Delta(I), l\in L\}$ such that, at stage $n \in
\mathbb{N}$, Player~1 chooses  a type $l_n$ and, given this choice,
$i_n$ is drawn accordingly to $x(l_n)$. We assume that $L$ is finite
since otherwise Player~1 have trivial strategies that guarantee that
the frequency of every $l$ converges to zero. Note that since the
choices of $l_n$ can be random, any behavioral strategy can be
described in such a way.

\begin{definition}[Lehrer-Solan~\cite{LehrerSolanPSE}]\label{defMICexistence}

For every $n \in \mathbb{N}$ and every $l \in L$, the average
internal regret of type $l$ at stage $n$ is
\[\mathcal{R}_n(l)=\sup_{x\in\Delta(I)}\left[G(x,\overline{\mu}_n(l))-G(\overline{\imath}_n(l),\overline{\mu}_n(l))\right].\]

A strategy $\sigma$ of Player~1 is $(L,\varepsilon)$-internally
consistent  if for every strategy $\tau$ of Player~2:
\[ \limsup_{n \to +\infty} \frac{|N_n(l)|}{n}\bigg(\mathcal{R}_n(l) - \varepsilon\bigg) \leq 0, \quad \forall l \in L, \quad \mathbb{P}_{\sigma,\tau}\mbox{-as}.\]
\end{definition}

\begin{remark}
Note that this definition, unlike in the full monitoring case, is
not intrinsic. It depends on the choice (which can be assumed to be
made by Player~1) of $\{x(l), l \in L\}$, and is based uniquely on
the potential observations (i.e.\ the sequences of flags $(\mu_n)_{n
\in \mathbb{N}}$) of Player~1.
\end{remark}
\begin{remark}
The average flag $\overline{\mu}_n$ belongs to $\Delta(S)^I$ and is
defined by $\overline{\mu}^i_n[s]=\frac{\sum_{m=1}^n \mu_m^i[s]}{n}$
 for  every  $s \in S$.
\end{remark}

In order to construct $(L,\varepsilon)$-internally consistent
strategies, some regularity over $G$ is required:
\begin{hypo}\label{hypo01}
For every $\varepsilon >0$, there exist  $\big\{\mu(l) \in
\Delta(S)^I, x(l) \in \Delta(I), l \in L\big\}$ two finite families
and  $\eta, \delta>0$  such that:
\begin{enumerate}
 \item{$\Delta(S)^I \subset \bigcup_{l \in L}B(\mu(l),\delta)$;}
\item{For every $l \in L$, if $\left\| x-x(l)\right\| \leq 2\eta$ and $\left\| \mu-\mu(l)\right\|\leq 2\delta$, then $x \in BR_{\varepsilon}(\mu)$,}
\end{enumerate}
where $BR_{\varepsilon}(\mu)=\left\{x \in \Delta(I): G(x,\mu) \geq
\sup_{z \in \Delta(I)}G(z,\mu)-\varepsilon \right\}$ is the set of
$\varepsilon$-best
 response to $\mu \in \Delta(S)^I$ and $B(\mu,\delta)=\left\{\mu' \in \Delta(S)^I, \left\| \mu'-\mu\right\| \leq \delta\right\}$.
\end{hypo}

In words, Assumption \ref{hypo01} implies that $G$ is regular with
respect to $\mu$ and with respect to $x$: given $\varepsilon$, the
set of flags can be covered by a finite number of balls centered in
$\{\mu(l), l \in L\}$, such that $x(l)$ is an $\varepsilon$-best
response to any $\mu$ in this ball. And if $x$ is close enough to
$x(l)$, then $x$ is also an $\varepsilon$-best response to any $\mu$
close to $\mu(l)$. Without loss of generality, we can assume that
$x(l)$ is different from $x(l')$ for any $l \neq l'$.

\begin{theorem}\label{MICexistence1}
Under Assumption \ref{hypo01}, there exist
$(L,\varepsilon)$-internally consistent strategies.
\end{theorem}
Some parts of the proof are quite technical, however the insight is
very simple,  so we give  firstly the main ideas. Assume for the
moment that Player~1 fully observes the flag at each stage. If, in
the one stage game, Player~2 plays first and his choice generates a
flag  $\mu \in \Delta(S)^I$, then Player~1 has an action  $x \in
\Delta(I)$ such that $x$ belongs to $BR_{\epsilon}(\mu)$. Using a
minmax argument (like Blackwell did for the proof of Theorem
\ref{repoussable2}, recall Remark~\ref{remarkapproch} $b)$  one
could prove that Player~1 has an $(L,\varepsilon)$-internally
consistent strategy (as did Lehrer and Solan~\cite{LehrerSolanPSE}).

\bigskip

The idea is to use calibration  to transform this implicit proof
into a constructive one, as in the alternative proof of Corollary
\ref{repoussable2}. Fix $\varepsilon>0$  and consider the game where
Player~1 predicts the sequence $(\mu_n)_{n \in \mathbb{N}}$ using
the $\delta$-grid $\{\mu(l),l\in L\}$ given by Assumption
\ref{hypo01}. A calibrated
 strategy of Player~1 chooses a sequences $(l_n)_{n \in \mathbb{N}}$
in such a way that $\overline{\mu}_n(l)$ is asymptotically
$\delta$-close to $\mu(l)$. Hence Player~1 just has to play
accordingly to $x(l)\in BR_{\varepsilon}(\mu(l))$ on these stages.

Indeed, since the choices of action are independent,
$\overline{\imath}_n(l)$ will be asymptotically $\eta$-close to
$x(l)$ and the regularity of $G$  will imply then that
$\overline{\imath}_n(l) \in BR_{\varepsilon}(\overline{\mu}_n(l))$
and so the strategy will be $(L,\varepsilon)$-internally consistent.

\medskip

The only issue is that in the current framework the signal depends
on the action of Player~1  since the law of $s_n$ is the $i_n$
component of $\mu_n$, which is not observed. Signals (that belong to
$S$) and  predictions (that belong to  $\Delta(S)^I$) are in two
different spaces, so the existence of calibrated strategies is not
straightforward. However, it is well known that, up to a slight
perturbation of $x(l)$, the information available to Player~1 after
a long time is close to $\overline{\mu}_n(l)$ (as in the multi-armed
bandit problem, some calibration and no-regret frameworks, see
e.g.\ Cesa-Bianchi and Lugosi~\cite{CesaBianchiLugosi} chapter 6 for a survey on these
techniques).

For every $x \in \Delta(I)$, define $x_{\eta} \in \Delta(I)$, the
$\eta$-perturbation of $x$ by $x_{\eta}=(1-\eta)x+\eta \mathbf{u}$
with $\mathbf{u}$ the uniform probability over $I$ and for every
$n$ define $\widehat{s}_n$ by:
\[\widehat{s}_n=\left(\frac{\mathbf{1}\{s_n=s\}\mathbf{1}\{i_n=i\}}{x_{\eta}(l_n)[i_n]}\right) \in \mathbb{R}^{SI},\]
with $x_{\eta}(l_n)[i_n]\geq \eta >0$ the weight put by $x_{\eta}(l_n)$ on $i_n$. We denote by
 $\widetilde{s}_n(l)$, instead of $\overline{\widehat{s}}_n(l)$, their average on $N_n(l)$.
\begin{lemma}\label{infomax}
For every $\theta>0$, there exists $N\in \mathbb{N}$ such that, for
every $l \in L$:
\[\mathbb{P}_{\sigma,\tau}\left.\left(\forall m \geq n, \left\| \widetilde{s}_n(l)-\overline{\mu}_n(l)\right\|\leq\theta\right|N_n(l) \geq
N\right)\geq1-\theta.\]
\end{lemma}

\textbf{Proof: }   Since for every $n \in \mathbb{N}$, the choices
of $i_n$ and $\mu_n$ are independent:
\begin{eqnarray}\nonumber \mathbb{E}_{\sigma,\tau}\left[\left.\widehat{s}_n\right|h_{n-1},l_n,\mu_n\right]&=&\sum_{i \in I}\sum_{s\in
S}  \mu_n^{i}[s]
x_{\eta}(l_n)[i]\left(0,\dots,\frac{s}{x_{\eta}(l_n)[i]},\dots,0\right)\\
\nonumber&=&\sum_{i \in I}\sum_{s \in
S}\mu_n^i[s]\left(0,\dots,s,\dots,0\right)\\ \nonumber& =&\sum_{i
\in
I}\left(0,\dots,\mu_n^{i},\dots,0\right)\\
\nonumber&=&\left(\mu_n^1,\dots,\mu_n^I\right)=\mu_n,\end{eqnarray}
where $\mu_n$ is seen as an element of $\mathbb{R}^{SI}$. Therefore
$\widetilde{s}_n(l)$ is an unbiased estimator of
$\overline{\mu}_n(l)$ and Hoeffding-Azuma's inequality (actually its
multidimensionnal version by Chen and White~\cite{ChenWhite}
together with the fact that $\sup_{n \in
\mathbb{N}}\|\widehat{s}_n\| \leq \eta^{-1}<\infty$) implies that
for every $\theta
>0$ there exists $N \in \mathbb{N}$ such that, for every $l \in L$:
\[ \mathbb{P}_{\sigma,\tau}\left.\left(\forall m \geq n, \left\| \widetilde{s}_n(l)-\overline{\mu}_n(l)\right\|\leq\theta\right||N_n(l)| \geq
N\right)\geq1-\theta. \] $\hfill \Box$

Assume now that Player~1 uses a calibrated strategy to predict the
sequences of $\widehat{s}_n$ (this is game is in full monitoring),
then he knows that asymptotically $\widetilde{s}_n(l)$ is closer to
$\mu(l)$ than to any $\mu(k)$ (as soon as the frequency of $l$ is
big enough), therefore it is $\delta$-close to $\mu(l)$. Lemma
\ref{infomax} implies that $\overline{\mu}_n(l)$ is asymptotically
close to $\widetilde{s}_n(l)$ and therefore $2\delta$-close to
$\mu(l)$.

\bigskip
\textbf{Proof of Theorem \ref{MICexistence1}:} Let the families
$\{x(l) \in \Delta(I), \mu(l) \in \Delta(S)^I,l\in L\}$ and
$\eta,\delta>0$ be given by
 Assumption \ref{hypo01} for a fixed $\varepsilon>0$.

\bigskip
Let $\Gamma'$ be the auxiliary repeated game where, at stage $n$,
Player~1 (resp.\ Player~2) chooses $l_n \in L$ (resp.\ $\mu_n \in
\Delta(S)^I$). Given these choices, $i_n$ (resp.\ $s_n$) is drawn
accordingly to $x_{\eta}(l_n)$ (resp.\ $\mu_n^{i_n}$). By Lemma
\ref{infomax}, for every $\theta >0$, there exists $N_1 \in
\mathbb{N}$ such that for every $l \in L$:
\begin{equation}\label{prooflemme}\mathbb{P}_{\sigma,\tau}\left.\left(\forall m \geq n,
\left\|
\widetilde{s}_n(l)-\overline{\mu}_n(l)\right\|\leq\theta\right||N_n(l)|
\geq N_1\right)\geq1-\theta.\end{equation} Let $\sigma$ be a
calibrated strategy associated to $(\widetilde{s}_n)_{n \in
\mathbb{N}}$ in $\Gamma'$. For every $\theta>0$, there exists $N_2
\in \mathbb{N}$ such that  with
$\mathbb{P}_{\sigma,\tau}$-probability greater than $1-\theta$:
\begin{equation}\label{proofcalib}\forall n \geq N_2,
\forall l,k \in L,
\frac{|N_n(l)|}{n}\bigg(\left\|\widetilde{s}_n(l)-\mu(l)\right\|^2-\left\|\widetilde{s}_n(l)-\mu(k)\right\|^2\bigg)\leq
\theta.\end{equation} Since $\{\mu(k), k \in L\}$ is a $\delta$-grid
of $\Delta(S)^I$, for every $n \in \mathbb{N}$ and $l \in L$, there
exists $k \in L$ such that $\left\|\widetilde{s}_n(l)-\mu(k)\right\|
\leq \delta$. Therefore, combining equation (\ref{prooflemme}) and
(\ref{proofcalib}), for every $\theta >0$ there exists $N_3 \in
\mathbb{N}$ such that:
\begin{equation}\label{fin1} \mathbb{P}_{\sigma,\tau}\left(\forall n \geq
N_3, \forall l \in L,
\frac{|N_n(l)|}{n}\bigg(\left\|\overline{\mu}_n(l)-\mu(l)\right\|^2-\delta^2\bigg)\leq\theta,
\right)\geq 1-\theta.\end{equation}

For every stage of type $l \in L$, $i_n$ is drawn accordingly to
$x_{\eta}(l)$ and by definition $\left\| x_{\eta}(l)-x(l)\right\|
\leq \eta$. Therefore Hoeffding-Azuma's inequality implies that, for
every $\theta >0$ there exists $N_4 \in \mathbb{N}$ such that:
\begin{equation}\label{fin3} \mathbb{P}_{\sigma,\tau}\left(\forall n \geq N_4, \forall l
\in L,
\frac{|N_n(l)|}{n}\bigg(\left\|\overline{\imath}_n(l)-x(l)\right\|-\eta\bigg)\leq\theta,
\right)\geq 1-\theta.\end{equation} Combining equation (\ref{fin1}),
(\ref{fin3}) and using Assumption \ref{hypo01}, for every $\theta
>0$, there exists $N \in \mathbb{N}$ such that for every strategy
$\tau$ of Player~2:
\begin{equation}\label{fin2} \mathbb{P}_{\sigma,\tau}\left(\forall n \geq N, \forall l
\in L,
\frac{|N_n(l)|}{n}\bigg(\mathcal{R}_n(l)-\varepsilon\bigg)\leq\theta,
\right)\geq 1-\theta,\end{equation} and $\sigma$ is
$(L,\varepsilon)$-internally consistent. $\hfill \Box$

\begin{remark}\label{remarkrate}
Lugosi, Mannor and Stoltz~\cite{LugosiMannorStoltz} provided an
algorithm that constructs, by block of size $m \in \mathbb{N}$, a
strategy that has  no external regret. We can describe it as
follows. Play at every stage of the $k$-th block $B_k$  according to
the same probability $x_k \in \Delta(I)$. Then compute (using Lemma
\ref{infomax}) an estimator of the average flag on this bloc and
denote it by $\widetilde{\mu}_k$. Knowing this flag, compute the
average regret accumulated on this specific block and aggregate it
to the previous regret in order to estimate the average regret from
the beginning of the game. Decide next what action is going to be
played on the following block according to a classical exponential
weight algorithm. With a fine tuning of $m \in \mathbb{N}$ (and
$\eta>0$), the external regret of this strategy converges to zero at
the rate  $O\left(n^{-1/5}\right)$ (the optimal rate is known to be
$n^{-1/3}$).

\medskip

Instead of trying to compute (or at least approximate) the sequence
of payoffs from the sequence of signals, our algorithm consider an
abstract auxiliary game defined on the signal space (i.e.\ on the
relevant information, the observations). We define payoffs in this
abstract game in order to transform it into a game with full
monitoring: the action set of Player~2 are flags, that are (almost)
observed by Player~1.

\bigskip

The strategy constructed is based on $\delta$-calibration and
Hoeffding-Azuma's inequality, therefore one can show that:
\[
\mathbb{E}_{\sigma,\tau}\left[\sup_{l \in
L}\frac{|N_n(l)|}{n}\bigg(\mathcal{R}_n(l)-\varepsilon\bigg)\right]\leq
O\left(\frac{1}{\sqrt{n}}\right).
\]
So given $\varepsilon>0$, one can construct a strategy such that the
internal regret converges quickly to $\varepsilon$, but maybe very
slowly to zero (because the constants depend, once again, drastically on
$\varepsilon^{J}$).
\end{remark}

\begin{remark}
Since  $\widetilde{s}_n$  converges to $\overline{\mu}_n$,  the
regret can be defined in terms of observed empirical flags instead
of unobserved average flag. For the same reason, $x(l)$ can be used to  define
regret.
\end{remark}
\subsection{On the strategy space}
One might object that behavioral strategies of Players 1 are defined
as mappings from the set of past histories $H^1=\bigcup_{n \in
\mathbb{N}}\left(I \times S\right)^n$ into $\Delta(I)$ while in
Definition \ref{defMICexistence} (and Theorem \ref{MICexistence1})
strategies considered are defined  as  mappings from $\bigcup_{n \in
\mathbb{N}}\left(I \times S \times L\right)^n$ into $\Delta(L)$,
with the specification that given $l_n \in L$, the law of $i_n$ is
$x(l_n)$ --- for a fixed family $\{x(l), l\in L\}$. Hence, they can
be defined as mappings from $\bigcup_{n \in \mathbb{N}}
\left(X\times I \times S\right)^n$ into $\Delta(X)$ (where
$X=\Delta(I)$ and $\Delta(X)$ is embedded with the star-weak
topology) and thus are behavioral  strategies in the game  where
Player~1's action set is $X$ and he receives at each stage a signal
in $I \times S$.

Therefore, they are  equivalent  to (i.e., following   Mertens Sorin
and Zamir~\cite{MSZ}, Theorem 1.8 p.\ 55, generate the same
probability on the set of plays as) mixed strategies, which are
mixtures of pure strategies, i.e.\ mappings from $\bigcup_{n \in
\mathbb{N}} \left(X \times I \times S\right)^n$ into $X$. These
latter are equivalent to  applications from $\bigcup_{n \in
\mathbb{N}} \left(I \times S\right)^n$ into $X$. Indeed, consider
for example $\sigma: \bigcup_{n \in \mathbb{N}}\left(X \times
T\right)^n\to X$ and define $\widetilde{\sigma}: \bigcup_{n \in
\mathbb{N}} T^n\to X$ recursively by
$\widetilde{\sigma}(\emptyset)=\sigma(\emptyset)$ and
\[\widetilde{\sigma}\left(t_1,\dots,t_n\right)=\sigma\left(\widetilde{\sigma}(\emptyset),t_0,\dots,\widetilde{\sigma}(t_0,\dots,t_{n-1}),t_n\right).\]

Finally, they are, in the game where
Player~1's action set is $I$ and he receives at each stage a signal
in $S$, mixtures of behavioral strategies --- also called general strategies --- so
 are equivalent to  behavioral strategies.

In conclusion, given a strategy defined as in Definition \ref{defMICexistence}, there exists a
behaviorial strategy that generates the same probability on the
set of plays (for every strategy $\tau$ of Player~2).

In these general strategies, Player~1 uses two types of signals: the
signals generated by \textsl{the game}, i.e.\ the sequence
$(i_n,s_n)_{n \in \mathbb{N}}$ and some private signals generated by
\textsl{his own strategy}, i.e.\ the sequences of $l_n$. We can
compute internal regret in Theorem \ref{MICexistence1} not only
because  the choices of $\mu_n$ and $l_n$ are independent given the
past, but mainly because the choices of $\mu_n$ and $i_n$ are
independent, even when $l_n$ is known.

\section{Back on payoff space}\label{sectionexample}
In the section we give simple condition on $G$ that ensures it fulfills  Assumption \ref{hypo01}. We also extend the framework to the so-called \textsl{compact case}. Finally, we prove that an internally consistent strategy (in a sense to be specified) is also externally consistent.

\subsection{The worst case fulfills Assumption \ref{hypo01}}
\begin{proposition}\label{hypo00}
Let $G: \Delta(I) \times \Delta(S)^I$ be such that for every $\mu
\in \Delta(S)^I$, $G(\cdot,\mu)$ is continuous and the family
$\{G(x,\cdot), x \in \Delta(I)\}$ is equicontinuous.

Then $G$ fulfills  Assumption \ref{hypo01}.
\end{proposition}
\textbf{Proof: } Since $\{G(x,\cdot), x \in \Delta(I)\}$ is
equicontinuous and $\Delta(S)^I$ compact, for every $\varepsilon>0$,
there exists $\delta>0$ such that:
\[ \forall x \in \Delta(I), \forall \mu,\mu' \in \Delta(S)^I, \| \mu - \mu'\| \leq 2\delta \Rightarrow \left|G(x,\mu)-G(x,\mu')\right| \leq \frac{\varepsilon}{2}.\]
Let $\{\mu(l), l \in L\}$ be a finite $\delta$-grid of $\Delta(S)^I$
 and  for every $l \in L$,
$x(l) \in BR(\mu(l))$ so that $G(x(l),\mu(l))=\max_{z \in
\Delta(I)}G(z,\mu(l))$. Since $G(x(l),\cdot)$ is continuous, there
exists $\eta(l)
>0$ such that:
\[\left\| x-x(l)\right\| \leq \eta(l)\Rightarrow\left|G(x,\mu(l))-G(x(l),\mu(l))\right| \leq \varepsilon/2.\]

\medskip

Define $\eta=\min_{l \in L} \eta(l)$ and let $x \in \Delta(I)$, $\mu
\in \Delta(S)^I$ and $l \in L$ such that $\left\| x-x(l)\right\|
\leq \eta$ and $\left\| \mu-\mu(l)\right\| \leq \delta$, then:
\[ G(x,\mu)\geq G(x,\mu(l))-\frac{\varepsilon}{2} \geq G(x(l),\mu(l))-\varepsilon= \max_{z \in \Delta(I)}G(z,\mu(l))-\varepsilon,\]
and $x \in BR_{\varepsilon}(\mu)$.  $\hfill \Box$

\medskip

This proposition implies that the evaluation function used by
Rustichini fulfills  Assumption \ref{hypo01} (see also Lugosi, Mannor
and Stoltz~\cite{LugosiMannorStoltz}, Lemma 3.1 and Proposition A.1).
Before proving that, we introduce $\mathcal{S}$, the range of
$\mathbf{s}$, which is a closed convex subset of $\Delta(S)^I$,  and
$\Pi_{\mathcal{S}}(\cdot)$ the projection onto it.
\begin{corollary}\label{corolworst}
Define $W: \Delta(I)\times \Delta(S)^I \to \mathbb{R}$ by:
\[ W(x,\mu) = \left\{\begin{array}{cc}\inf_{y \in \mathbf{s}^{-1}(\mu)}\rho(x,y) & \mbox{if } \quad \mu \in \mathcal{S}\\W\left(x,\Pi_{\mathcal{S}}(\mu)\right)& \mbox{otherwise.}\end{array}\right.
\]
Then $W$ fulfills Assumption \ref{hypo01}.
\end{corollary}
\textbf{Proof: } We extend  $\mathbf{s}$ linearly to
$\mathbb{R}^{J}$ by $\mathbf{s}(y)= \sum_{j \in J}y(j)\mathbf{s}(j)$
where $y=(y(j))_{j \in J}$. Therefore (Aubin and Frankowska
\cite{AubinFrankowska}, Theorem 2.2.1, p.\ 57) the multivalued
application $\mathbf{s}^{-1}: \mathcal{S} \rightrightarrows
\Delta(J)^I$ is $\lambda$-Lipschitz, and since $\Pi_{\mathcal{S}}$
is 1-Lipschitz (because $\mathcal{S}$ is convex), $W(x, \cdot)$ is
also $\lambda$-Lipschitz, for every $x \in \Delta(I)$. Therefore,
$\{G(x,\cdot),x \in \Delta(I)\}$ is equicontinuous. For every $\mu
\in \Delta(S)^I$, $W(\cdot, \mu)$ is $r$-Lipschitz (where
$r=\|\rho\|$, see e.g.\ Lugosi, Mannor and Stoltz
\cite{LugosiMannorStoltz}), therefore continuous. Hence, by
Proposition \ref{hypo00}, $W$ fulfills Assumption \ref{hypo01}.
$\hfill \Box$
\subsection{Compact case}
Assumption \ref{hypo01} does not require that Player~1 faces only one
opponent, nor that his opponents have only a finite set of actions.
As long as $G$ is regular then Player~1 has a
$(L,\varepsilon)$-internally consistent strategy, for every
$\varepsilon>0$. We consider in this section a particular framework,
referred as the {\em compact case} (as mentioned in section \ref{sectionfull}).

Player~1's action set is still denoted by $I$, but we now assume
that the action set of Player~2 is $[-1,1]^I$. The payoff mapping
$\rho$ from $\Delta(I) \times [-1,1]^I$ to $\mathbb{R}$ is simply
defined by $\rho(x,U)=\langle x, U \rangle$. Let $\mathbf{s}$ be a
multivalued application from $[-1,1]^I$ to $\Delta(S)^I$. Given the
choices of $i$ and $U$, Player~1 does not observe $U$ but receives a
signal $s \in S$, whose law is the $i$-th component of $\mu$ which
belongs to $\mathbf{s}(U)$. If $\mathbf{s}(U)$  is  not  a singleton
then we can assume either that $\mu$ is chosen by Nature (a third
player) or by Player~2.

\bigskip
A multivalued application $\mathbf{s}$ is closed-convex if $ \lambda
\mathbf{s}(x)+(1-\lambda)\mathbf{s}(z)\subset \mathbf{s}(\lambda
x+(1-\lambda)z)$ and its graph is closed and its inverse is defined by $\mathbf{s}^{-1}(\mu)=\{U \in [-1,1]^I,
\mu \in \mathbf{s}(U)\}$. It is clear that if $\mathbf{s}$ is
closed-convex then $\mathbf{s}^{-1}$ is also closed-convex.
\begin{proposition}
Define the worst case mapping as in Corollary \ref{corolworst}. If
$\mathbf{s}$ is closed-convex and its range is a polytope (the
convex hull of a finite number of points), then $W$ fulfills
Assumption \ref{hypo01}.
\end{proposition}
\textbf{Proof: } We follow Aubin et
Frankowska~\cite{AubinFrankowska}: let $\mu_0$ be in $\mathcal{S}$
the range of $\mathbf{s}$,  $U_0$ be in $\mathbf{s}^{-1}(\mu_0)$ and
$g$ be the mapping defined by:
\begin{eqnarray}
\nonumber g: \mathcal{S} & \mapsto& \mathbb{R}\\
\nonumber \mu & \to & g(\mu)=\inf_{U \in
\mathbf{s}^{-1}(\mu)}\left\|U-U_0\right\|=d\left(U_0,\mathbf{s}^{-1}(\mu)\right).
\end{eqnarray}
Since $\mathbf{s}$ is convex, so is $\mathbf{s}^{-1}$ (in the
multivalued sense) and $g$ (in the univalued sense). The sections
$\{\mu| g(\mu) \leq \lambda\}$ are closed (see
Aubin and Frankowska~\cite{AubinFrankowska}, Lemma 2.2.3 p.\~59) so $g$ is lower
semi-continuous. Since the domain of $g$ is a polytope, $g$ is
also upper semi-continuous (see Rockafellar~\cite{Rockafellar}, Theorem 10.2
p.\ 84). Therefore $g$ is continuous over $\mathcal{S}$ and there
exists $\delta(U_0)$ such that if $\|\mu-\mu_0\|\leq \delta(U_0)$
then $d\left(U_0,\mathbf{s}^{-1}(\mu)\right)\leq \varepsilon$.

Since $\mathbf{s}^{-1}(\mu_0)$ is compact, for every
$\varepsilon >0$, there exists a finite set $\mathcal{U}$ such that $\mathbf{s}^{-1}(\mu_0) \subset \bigcup_{U \in \mathcal{U}}B(U,\varepsilon)$. Define $\delta(\mu_0)=\inf_{U \in \mathcal{U}}\delta(U_0)$, then for every $\mu$ in $\Delta(S)^I$,
$\|\mu-\mu_0\|\leq \delta(\mu_0)$ implies that
$\mathbf{s}^{-1}(\mu_0)\subset \mathbf{s}^{-1}(\mu)+2\varepsilon B$
(with $B$ the unit ball). The graph of $\mathbf{s}^{-1}$ is compact
so for every $\varepsilon>0$ there exists
$0<\delta'(\mu_0)<\delta(\mu_0)$ such that if $\|\mu-\mu_0\|\leq
\delta'(\mu_0)$ then $\mathbf{s}^{-1}(\mu)\subset
\mathbf{s}^{-1}(\mu_0)+2\varepsilon B$.

There exists a finite set $M$ such that the compact set
$\mathcal{S}$ is included in the union of open balls $\bigcup_{\mu \in M}B(\mu, \delta'(\mu)/3)$. If we denote
by $\delta=\inf_{\mu \in M}\delta'(\mu)/3$ then for every $\mu$ and
$\mu'$ in $\mathcal{S}$, if $\|\mu-\mu'\|\leq \delta$, there exists
$\mu_1 \in M$ such that $\mu$ and $\mu'$ belongs to
$B(\mu_1,\delta'(\mu_1))$ hence $\mathbf{s}^{-1}(\mu) \subset
\mathbf{s}^{-1}(\mu_1)+2\varepsilon B \subset
\mathbf{s}^{-1}(\mu')+4\varepsilon B$.

Let $\mu$ and $\mu'$ in $\Delta(S)^I$ such that $\|\mu-\mu'\| \leq
\delta$. Then since $\mathcal{S}$ is a convex set
$\|\Pi_{\mathcal{S}}(\mu)-\Pi_{\mathcal{S}}(\mu')\| \leq \delta$ and for every $x \in \Delta(I)$
\[W(x,\mu)=\inf_{U \in \mathbf{s}^{-1}(\Pi_{\mathcal{S}}(\mu))}\langle x, U \rangle \geq \inf_{U \in \mathbf{s}^{-1}(\Pi_{\mathcal{S}}(\mu'))}\langle x, U \rangle -4\varepsilon=W(x,\mu')-4\varepsilon.\]
Let $x$ and $x'$ in $\Delta(I)$ such that $\|x-x'\|\leq \varepsilon$
then for all $\mu \in \Delta(S)^I$
\[W(x,\mu)=\inf_{U \in \mathbf{s}^{-1}(\Pi_{\mathcal{S}}(\mu))}\langle x, U \rangle \geq \inf_{U \in \mathbf{s}^{-1}(\Pi_{\mathcal{S}}(\mu))}\langle
x', U \rangle -\varepsilon=W(x',\mu)-\varepsilon.\] Hence if $x(l)$ is a $\varepsilon$-best response
to $\mu(l)$, $\|x-x(l)\|\leq\varepsilon$ and $\|\mu-\mu(l)\|\leq
\delta$ then
\begin{eqnarray}\nonumber W(x,\mu)\geq W(x(l),\mu)-\varepsilon\geq W(x(l),\mu(l))-5\varepsilon&\geq& \sup_{z \in \Delta(I)}W(z,\mu(l))-6\varepsilon\\
\nonumber &\geq & \sup_{z \in
\Delta(I)}W(z,\mu)-10\varepsilon,\end{eqnarray} so $x$ is a
$10\varepsilon$-best response to $\mu$. $\hfill \Box$

\begin{remark}[On the assumptions over $\mathbf{s}$]

$\mathbf{s}$ is assumed to be multivalued since in the finite case,
there might be two different probabilities $y$ and $y'$ in
$\Delta(J)$ that generate the same outcome vector
$\rho(y)=(\rho(i,y))_{i \in I}=\rho(y')$ but two different flags
$\mathbf{s}(y)$ and $\mathbf{s}(y')$.

It is also convex: if Player~2 can generate a flag $\mu$ by
playing $y \in \Delta(J)$ and a flag $\mu'$ by playing $y'$, then a
convex combination of $y$ and $y'$ should generate the same convex
combination of flags. This assumption is specifically needed with
repeated game: for example, Player~2 can play $y$ on odd stages and $y'$ on even
stages. Player~1 must know that the average empirical flag can
be generated by $1/2y+1/2y'$.

The fact that the range of $\mathbf{s}$ is a polytope (or at least
that it is locally simplicial, see Rockafellar~\cite{Rockafellar}
p.\ 84 for formal definitions and examples)  is needed for the proof
that $W$ is continuous.  It is obviously true in the finite
dimension case since its  graph is a polytope.
\end{remark}

\subsection{Regret in terms of actual payoffs}
As Rustichini~\cite{Rustichini}, we can define  regret  in term of
unobserved average payoff.
\begin{definition}
A strategy $\sigma$ of Player~1 is $(L,\varepsilon)$-internally
consistent with respect to the actual payoffs if for every $l\in L$:
\[ \limsup_{n \to +\infty} \frac{|N_n(l)|}{n}\left(\sup_{x\in\Delta(I)}\left[W(x,\overline{\mu}_n(l))-\overline{\rho}_n(l)\right]-\varepsilon\right) \leq 0, \quad \mathbb{P}_{\sigma,\tau}\mbox{-as}.\]
\end{definition}

\begin{proposition}\label{propactual}
For every $\varepsilon>0$, there exist $(L,\varepsilon)$-internally
consistent strategies with respect to the actual payoffs.
\end{proposition}
\textbf{Proof: }   Consider the strategy $\sigma$ given by Theorem
\ref{MICexistence1} with the worst case mapping. By definition of
$W$ and using the independence of the choices of $x(l)$ and $\mu_n$,
one can easily show that asymptotically
$W\left(x(l),\overline{\mu}_n(l)\right) \leq \overline{\rho}_n(l)$.
Therefore the strategy $\sigma$ is also $(L,\varepsilon)$-consistent
with respect to the actual payoffs. $\hfill \Box$

Now we can define $0$-internally consistent strategies (see Lehrer and Solan~\cite{LehrerSolanPSE} definition 10):
\begin{definition}\label{def0consist}
A strategy $\sigma$ of Player~1 is $0$-internally consistent if for
every $\varepsilon>0$, there exists $\delta>0$ such that for every
finite partition $\{P(l),l \in L\}$ of $\Delta(I)$ with diameter
smaller than $\delta$ and every $l\in L$:
\[ \limsup_{n \to +\infty} \frac{|N_n(l)|}{n}\left(\sup_{x\in\Delta(I)}\left[W(x,\overline{\mu}_n(l))-\overline{\rho}_n(l)\right]-\varepsilon\right) \leq 0, \quad \mathbb{P}_{\sigma,\tau}\mbox{-as},\]
where $N_n(l)=\{m  \leq n, x_n \in P(l)\}$ with $x_n$ the law
(that might be chosen at random by Player~1) of $i_n$ given the past
history and $\overline{\mu}_n(l)$ (resp. $\overline{\imath}_n(l)$)
is the average flag (resp. action of Player~1) on $N_n(l)$.
\end{definition}
\begin{proposition}\label{0consiste}
There exist $0$-internally consistent strategies with respect to the
actual payoffs.
\end{proposition}
\textbf{Proof: }   The proof relies uniquely on a classical doubling
trick (see e.g.\ Sorin~\cite{SorinSupergames}, Proposition  3.2 p.\
56) recalled below.

Denote by $\sigma_k$ the strategy given by Proposition
\ref{propactual} for $\varepsilon_k=2^{-(k+3)}$. Consider the
strategy $\sigma$ of player defined by block: on the first block of
length $N_1$, Player~1 plays accordingly to $\sigma_1$, then on the
second block of length $N_2$ accordingly to $\sigma_2$, and so on.
Formally, for $n$ such that $\sum_{k=1}^{p-1}N_k \leq n  \leq
\sum_{k=1}^{p}N_k$, $\sigma(h_n)=\sigma_p(h_n^p)$ where
$h_n^p=\left(i_m,l_m,s_m\right)_{m \in \{ \sum_{k=1}^{p-1}N_k,
\dots, n\}}$ is the partial history on the last block. Remark
\ref{remarkrate} implies that for every $p \in \mathbb{N}$ there
exists $M_p \in \mathbb{N}$ such that
\[\mathbb{E}_{\sigma,\tau}\left[\sup_{l \in
L}\frac{|N_n(l)|}{n}\bigg(\mathcal{R}_n(l)\bigg)\right]\leq
\frac{1}{2^{p+1}}.\] Let $(N_k)_{k \in \mathbb{N}}$ be a sequence
such that $\sum_{p=1}^{k-1}N_p=o(N_k)$ and $M_{k+1}=o(N_k)$ (where
$u_n=o(v_n)$ means that $v_n>0$ and $\lim_{n \to
\infty}\frac{u_n}{v_n}=0$). With this definition,  the $m$-th block
is way longer than all the previous blocks, and longer than the time
required by $\sigma_{k+1}$ to be $\varepsilon_{k+1}$-consistent (in
expectation). So the (maybe high) regret accumulated during the
first $M_n$ stages of the $n$-th block  is negligible compared to
the small regret accumulated before (during the first
$(n-1)$-blocks). After these $M_n$ stages, the regret (on the $n$-th
block) is smaller than $\varepsilon_n$ and at the end of this block,
the cumulative regret is very close to $\varepsilon$.$\hfill \Box$

\begin{remark}
The use of a doubling trick prevents us to easily  find a bound on
the rate of convergence of the regret. The proof of Proposition
\ref{0consiste} requires that the sum of the regret on two different
block is smaller than the average regret. This is why we restrict
this definition to  internally consistent strategies with respect to
the actual payoffs. One may compare Definition \ref{def0consist} of
0-consistency to the Definition \ref{defcalib1} of
$\varepsilon$-calibrated strategies.\end{remark}

\subsection{External and internal consistency}
With full monitoring, by linearity of the payoff function, a
strategy that is internally consistent is also externally
consistent. This properties holds in partial monitoring, when we
consider regret in terms of actual payoffs:
\begin{proposition}\label{intetext}
For every $\varepsilon>0$ and $\{x(l), l \in L\}$ of $\Delta(I)$,
every $(L,\varepsilon)$-internally consistent strategy with respect
to the actual payoffs is $\varepsilon$-externally consistent with
respect to the actual payoffs, i.e.\ $\mathbb{P}_{\sigma,\tau}$-ps:
\[\limsup_{n \to +\infty}\max_{x \in \Delta(I)} W(x,\overline{\mu}_n)-\overline{\rho}_n\leq \varepsilon.\]\end{proposition}
\textbf{Proof: }   Let $\varepsilon>0$, $L \subset \Delta(I)$  and
$\sigma$ be  an $(L,\varepsilon)$-internally consistent strategy
with respect to the actual payoffs. Since $\mathbf{s}^{-1}(\cdot)$
is convex then, for every $x \in \Delta(I)$, the mapping $\mu
\mapsto W(x,\mu)$ is convex  and so is the mapping $\mu \mapsto
\max_{x \in \Delta(I)} W(x,\mu)$. Hence
\[\max_{x \in \Delta(I)}W(x,\overline{\mu}_n) - \overline{\rho}_n \leq \sum_{l \in L}\frac{|N_n(l)|}{n}\bigg(\max_{x \in
\Delta(I)}W(x,\overline{\mu}_n(l))-\overline{\rho}_n(l)\bigg).\]
Therefore, one has
\[\limsup_{n \to \infty} \max_{x \in \Delta(I)}W(x,\overline{\mu}_n) -
\overline{\rho}_n \leq \limsup_{n \to + \infty}\sum_{l \in
L}\frac{|N_n(l)|}{n}\varepsilon\leq \varepsilon\]and so $\sigma$ is
$\varepsilon$-externally consistent. $\hfill \Box$

\medskip
Proposition \ref{intetext} holds for the compact case under the
assumption that $\sigma$ is closed-convex. Note that the proof relies
on  the fact that $W$ is convex and the actual payoffs are linear. It
is clear that this result does not extend to any evaluation
function. Indeed, consider the optimistic function defined by (for
$\mu \in \mathcal{S}$):
\[O(x,\mu)=\sup_{y \in \mathbf{s}^{-1}(\mu)}\rho(x,y),\]
then  the more information about $\overline{\jmath}_n$ that Player~1
gets, the less he evaluates his payoff. So an internally consistent
strategy (i.e.\ a strategy that is consistent with a more precise
knowledge on the moves of Player~2) might not be externally
consistent.

\section*{Concluding remarks}
In the full monitoring framework, many improvements have been made
in the past years about calibration and regret (see for instance
\cite{lehrerWideRange,SandroniSmorodinskyVohra,Vovk}). Here, we aimed
to clarify the links between the original notions of
approachability, internal regret and calibration in order to extend
applications (in particular, to get rid of the finiteness of $J$),
to define the internal regret with signals as calibration over an
appropriate space and to give a proof derived from no-internal
regret in full monitoring, itself derived from the approachability
of an orthant in this space.

\bigskip

\textbf{Acknowledgments:} I deeply thanks my advisor Sylvain
Sorin for his great help and numerous comments. I also acknowledge
very helpful remarks from Gilles Stoltz.

\medskip

An extended abstract of this paper appeared in the \textsl{Proceedings of the 20th International Conference on Algorithmic Learning Theory}, Springer, 2009.


\begin{thebibliography}{99}
\bibitem{AubinFrankowska} J.-P. Aubin, H.  Frankowska,  Set-valued Analysis {\em Birkh\"auser Boston
Inc.} 1990

\bibitem{Azuma} K. Azuma,  {\em Weighted sums of certain dependent random variables}, T\^ohoku Math. J., vol.\ 19,
pp. 357--367, 1967


\bibitem{BlackwellAnalogue} D. Blackwell, {\em An analog of the minimax theorem for vector payoffs}, {\em Pacific J. Math.}, Vol. 6, pp.
1--8, 1956

\bibitem{BlackwellControlled}
D. Blackwell, {\em Controlled random walks}, {\em Proceedings of the
International Congress of Mathematicians},
              1954, Amsterdam, vol. III, pp. 336-338, 1956

\bibitem{CesaBianchiLugosi}
N. Cesa-Bianchi and G. Lugosi, {\em Prediction, Learning, and Games}
{\em Cambridge University Press}, 2006

\bibitem{CesaBianchiLugosiStoltz}
N. Cesa-Bianchi and G. Lugosi and G. Stoltz, {\em Minimizing regret
with label efficient prediction} {\em IEEE Trans. Inform. Theory},
vol. 51, pp. 2152--2162, 2005

\bibitem{ChenWhite} X. Chen and H. White, {\em Laws of large numbers for {H}ilbert space-valued mixingales
              with applications}, Econometric Theory, vol. 12, pp. 284--304, 1996


\bibitem{DawidWellCalibrated}
   A. P. Dawid, {\em The well-calibrated {B}ayesian},
  {\em J. Amer. Statist. Assoc.}, vol. 77, pp. 605--613, 1982

\bibitem{FosterVohraAsymptoticCalibration}    D. P . Foster and  R. V. Vohra,
{\em Asymptotic calibration} {\em Biometrika}, vol. 85, pp.
379--390, 1998


\bibitem{FosterVohraCalibratedLearningCorrelatedEquilibrium}
D. P. Foster  and R. V. Vohra, {\em Calibrated learning and
correlated equilibrium} {\em Games Econom. Behav.}, vol. 21,  pp.
40--515, 1997

\bibitem{FudenbergLevineConditionalUniversalConsistency}
D. Fudenberg and D. K. Levine, {\em Conditional universal
consistency} {\em Games Econom. Behav.}, vol. 29, pp. 104--130, 1999

\bibitem{Hannan} J. Hannan,  {\em Approximation to {B}ayes risk in repeated play}, {\em Contributions to the Theory of Games}, vol. 3,     97--139, 1957

\bibitem{HartMasColellCorrelatedEquilibrium}
S. Hart and A. Mas-Colell, {\em A simple adaptive procedure leading
to correlated equilibrium}, {\em Econometrica}, vol. 68, pp.
1127--1150, 2000

\bibitem{Hoeffding} W. Hoeffding,  {\em Probability inequalities for sums of bounded random variables},
  J. Amer. Statist. Assoc. Vol.\ 58, pp. 13--30, 1963

\bibitem{Khachiyan} L.G. Khachiyan, {\em Polynomial algorithms in linear
programming}, Zh. Vychisl. Mat. i Mat. Fiz., vol.\ 20, pp. 51-68,
1980

\bibitem{lehrerWideRange} E. Lehrer, {\em A wide range no-regret theorem},
  Games Econom. Behav., vol. 42, pp. 101--115, 2003


\bibitem{LehrerSolanPSE} E. Lehrer and E. Solan, {\em Learning to play partially-specified
equilibrium}, {\em manuscript}, 2007

\bibitem{LuceRaiffa} R.D. Luce and H. Raiffa, Games and Decisions: Introduction and Critical Survey, {\em John Wiley \& Sons Inc.}, 1957

\bibitem{LugosiMannorStoltz}
G. Lugosi and S. Mannor and G. Stoltz, {\em Strategies for
prediction under imperfect monitoring}, {\em Math. Oper. Res.}, vol.
33, pp. 513--528, 2008

\bibitem{MannorStoltz} S.  Mannor  and G. Stoltz, {\em A geometric proof of
calibration}, available at
http://hal.archives-ouvertes.fr/hal-00442042/fr/, 2010


\bibitem{MSZ} J.-F. Mertens and S. Sorin and S. Zamir,  Repeated Games {\em CORE discussion paper
9420-9422.} 1994


\bibitem{PerchetCalibrationALT}
  V. Perchet, {\em Calibration and Internal no-Regret with Random Signals}, {\em Proceedings of the 20th International Conference on Algorithmic
Learning Theory}, pp. 68--82, 2009

\bibitem{PerchetApproach}
V. Perchet {\em Approachability of convex sets in games with partial
monitoring},  manuscript, 2009

\bibitem{Rockafellar} R.T. Rockafellar, Convex Analysis,
{\em Princeton University Press}, 1970

\bibitem{Rustichini}
  A. Rustichini, {\em Minimizing regret: the general case}, {\em Games Econom.
  Behav.}, vol. 29, pp. 224--243, 1999


\bibitem{SandroniSmorodinskyVohra} A. Sandroni, R. Smorodinsky and R.V. Vohra, {\em Calibration with many checking rules},
 Math. Oper. Res., vol. 28, pp. 141--153, 2003

\bibitem{Seneta} E.  Seneta, Nonnegative Matrices and Markov Chains, {\em Springer Series in
Statistics}, 1981


\bibitem{SorinUnpublished} S. Sorin, Lectures on
Dynamics in Games,  {\em Unpublished Lecture Notes}, 2008

\bibitem{SorinSupergames} S. Sorin, {\em Supergames}, in: Game theory and applications, Academic Press, San Diego, CA, pp.
46--82, 1987

\bibitem{Vovk} V. Vovk, {\em Non-asymptotic calibration and resolution},
Theoret. Comput. Sci., vol. 387, pp. 77--89, 2007

\end{thebibliography}
\end{document}